\documentclass[prl,twocolumn,showpacs,superscriptaddress]{revtex4}
\usepackage{amssymb,amsmath}
\usepackage{graphicx,array}
\usepackage{dcolumn}
\usepackage{bm}

\begin{document}
\title{\emph{Ab initio} design of charge-mismatched ferroelectric superlattices}

\author{Claudio Cazorla}
\affiliation{Institut de Ci$\grave{e}$ncia de Materials de Barcelona
            (ICMAB-CSIC), 08193 Bellaterra, Spain}
\author{Massimiliano Stengel}
\affiliation{Institut de Ci$\grave{e}$ncia de Materials de Barcelona
            (ICMAB-CSIC), 08193 Bellaterra, Spain}
\affiliation{ICREA - Instituci\'o Catalana de Recerca i Estudis Avan\c{c}ats, 08010 Barcelona, Spain}
\email{ccazorla@icmab.es}

\begin{abstract}
We present a systematic approach to modeling the electrical and structural 
properties of charge-mismatched superlattices from first principles. 
Our strategy is based on bulk calculations of the parent compounds, which
we perform as a function of in-plane strain and out-of-plane electric
displacement field. 
The resulting two-dimensional phase diagrams allow us to 
accurately predict, without performing further calculations, 
the behavior of a layered heterostructure where the aforementioned
building blocks are electrostatically and elastically coupled, with
an arbitrary choice of the interface charge (originated from the polar 
discontinuity) and volume ratio.
By using the [PbTiO$_{3}$]$_{m}$/[BiFeO$_{3}$]$_{n}$ 
system as test case, we demonstrate that interface polarity has a dramatic
impact on the ferroelectric behavior of the superlattice, leading to
the stabilization of otherwise inaccessible bulk phases.
\end{abstract}

\maketitle

\section{Introduction}
\label{sec:intro}

\begin{figure}
\centerline{
\includegraphics[width=1.00\linewidth]{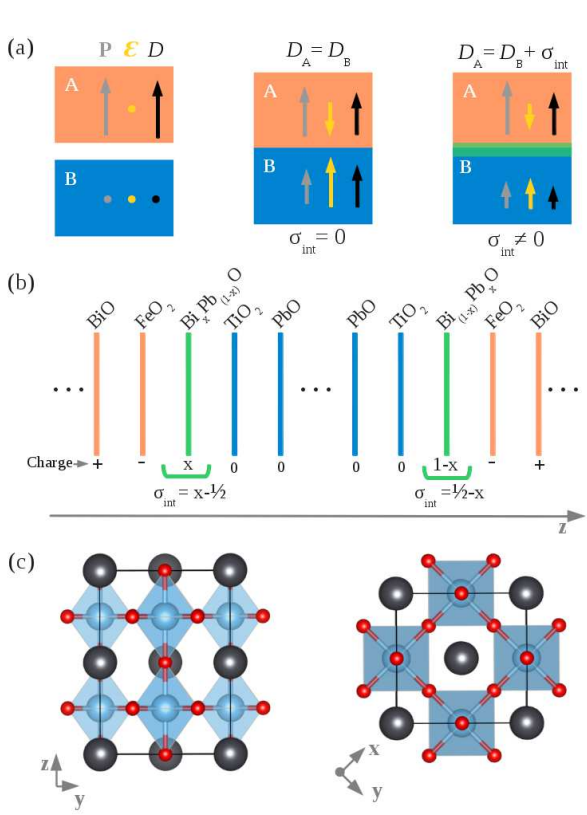}}
\vspace{-0.60cm}
\caption{(Color online) (a)~Description of the electrostatic coupling
         in a ferroelectric~(orange)/paraelectric~(blue) bilayer;
         $P$, $\mathcal{E}$, and $D$ represent the component
         of the polarization, electric field and electric
         displacement vectors along the stacking direction,
         and $\sigma_{\rm int}$ is the interface charge density.
         (b)~Intermixed AO-type interfaces in a
         [BiFeO$_{3}$]$_{m}$/[PbTiO$_{3}$]$_{n}$ superlattice
         and the resulting interface charge densities.
         (c)~Illustration of the $20$-atom simulation cell used in our calculations;
         red, blue and black spheres represent O, B, and A atoms in the
         ABO$_{3}$ perovskite.}
\label{fig1}
\end{figure}

When layers of perovskite oxides are epitaxially stacked to form a 
periodically repeated heterostructure, new intriguing functionalities
can emerge in the resulting superlattice~[\onlinecite{ghosez08,junquera11}]. 
These are further tunable via applied electric fields and thermodynamic
conditions, and thus attractive for nanoelectronics and energy applications. 
An excellent example is the [PbTiO$_{3}$]$_{m}$/[SrTiO$_{3}$]$_{n}$ 
system, where the polarization, tetragonality, piezoelectric response, 
and ferroelectric transition temperature strongly change 
with the volume ratio of the parent compounds~[\onlinecite{dawber05,dawber07,dawber12}]. 
Such a remarkable tunability is usually rationalized in terms of
epitaxial strains~[\onlinecite{dawber05b}], electrostatic coupling  
(see Fig.~\ref{fig1}a)~[\onlinecite{zubko12,wu12}], and local 
interface effects~[\onlinecite{junquera12,bousquet08}].

While perovskite titanates with ATiO$_3$ formula (A=Sr, Pb, Ba or Ca) have 
traditionally been the most popular choice as the basic building blocks, 
a much wider range of materials (e.g., BiFeO$_3$) is currently 
receiving increasing attention by the community. 
The motivation for such an interest is clear: a superlattice configuration
provides the unique opportunity of enhancing materials properties via 
``strain engineering'', and a multifunctional compound such as BiFeO$_3$
appears to be a natural candidate in this context. (For example, strain 
has been predicted to enhance the magnetoelectric response of BiFeO$_3$ by 
several orders of magnitude compared to bulk samples~[\onlinecite{wojdel09,wojdel10}].)
Also, a superlattice geometry can alleviate the leakage issues of
pure BiFeO$_3$ films~[\onlinecite{ranjith07,ranjith08}].
Combining a III--III perovskite like BiFeO$_3$ (or I--V, like KNbO$_3$) with a 
II--IV titanate appears, however, problematic from the conceptual point of view.
In fact, the charge-family mismatch inevitably leads to polar (and hence electrostatically 
unstable) interfaces between layers~\cite{murray09}.
This is not necessarily a drawback, though: recent research has demonstrated 
that polar interfaces can be, rather than a nuisance to be avoided, a rich 
playground to be exploited for exploring exciting new phenomena. 
The prototypical example is the 
LaAlO$_{3}$/SrTiO$_{3}$ system, where a metallic 
two-dimensional electron gas appears at the  
heterojunction to avoid a ``polar catastrophe''~\cite{nakagawa06,ohtomo04}.
Remarkably, first-principles calculations have shown that interfaces in 
oxide superlattices can remain insulating provided that the layers are 
thin enough, and produce rather dramatic effects on the respective polarization
of the individual components~\cite{bristowe09,murray09}.
This means that, in a superlattice, polar discontinuities need not be compensated 
by electronic or ionic reconstructions; they can, instead, 
be used as an additional, powerful materials-design tool to control the behavior of 
the polar degrees of freedom therein. Such a control may be realized, for instance, 
by altering the stoichiometry at the interfaces (see Fig.~\ref{fig1}b). 
To fully explore the potential that this additional degree of freedom (the interface
built-in polarity) provides, and guide the experimental search for the most
promising materials combinations, one clearly needs to establish a general
theoretical framework where the ``compositional charge''~\cite{murray09} is
adequately taken into account.

In this Letter, we present a general first-principles approach  
to predict the behavior of charge-mismatched perovskite oxide  
superlattices based exclusively on the properties 
of their individual bulk constituents. 
Our formalism combines the constrained-$D$ strategies of Wu {\em et al.}~\cite{wu08},
which are key to decomposing the total energy of the system into the contributions
of the individual layers, with the rigorous description of the interface
polarity proposed in Ref.~\cite{stengel11}.
As a result, we are able to exactly describe the electrostatic coupling and 
mechanical boundary conditions, enabling a clear 
separation between genuine interface and bulk effects. 
Crucially, the present method allows one to quantify, in a straightforward way, the 
impact that interface polarity has on the equilibrium 
(and metastable) phases of the superlattice. 
As a proof of concept we apply our formalism to the study of   
[PbTiO$_{3}$]$_{m}$/[BiFeO$_{3}$]$_{n}$ (PTO/BFO) heterostructures.
We find that (i)~our \emph{bulk} model accurately matches earlier first-principles 
predictions obtained for ultrashort-period superlattices 
(i.e., $m=n=3$) by using explicit \emph{supercell} simulations~\cite{stengel12},
and (ii)~by assuming interface terminations with different nominal charge,
we obtain a radical change in the overall ferroelectric properties of the
superlattice, which demonstrates the crucial role played by the polar mismatch.

\begin{figure}
\centerline
        {\includegraphics[width=1.00\linewidth]{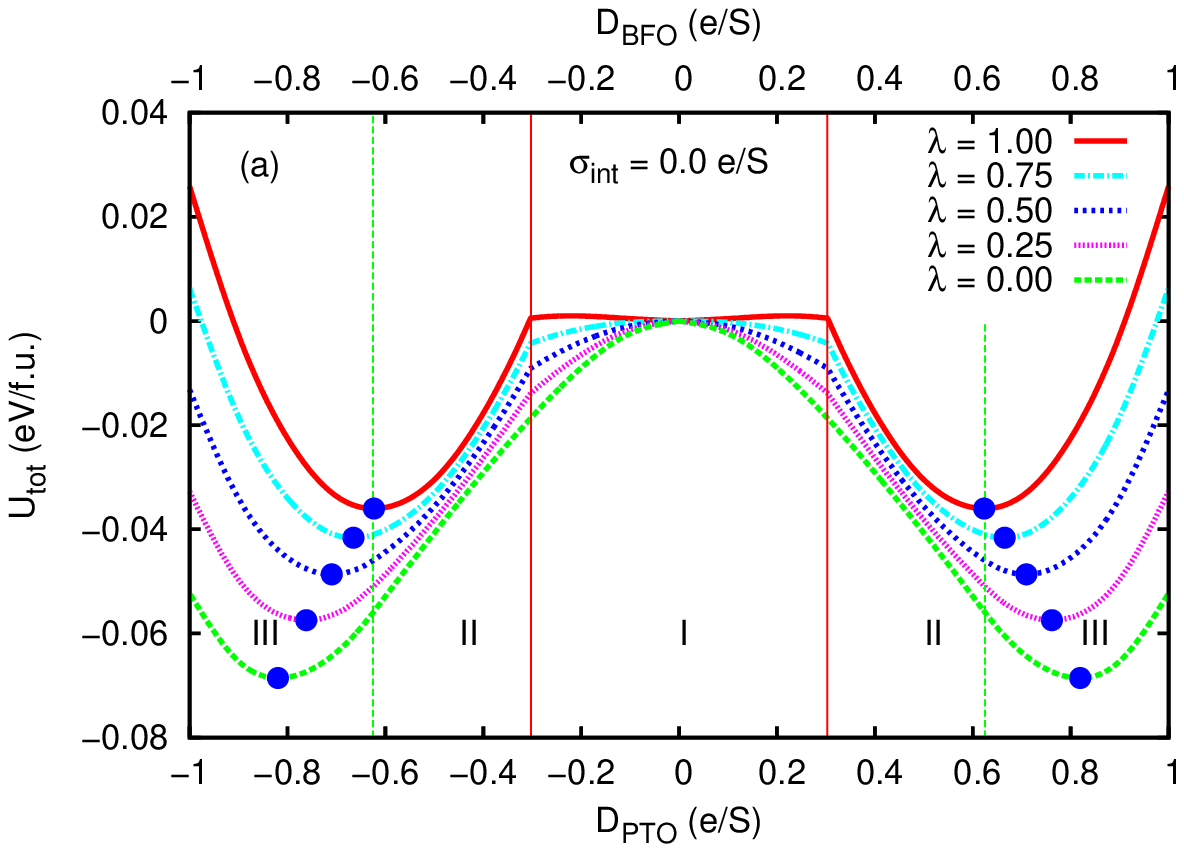}}%
        {\includegraphics[width=1.00\linewidth]{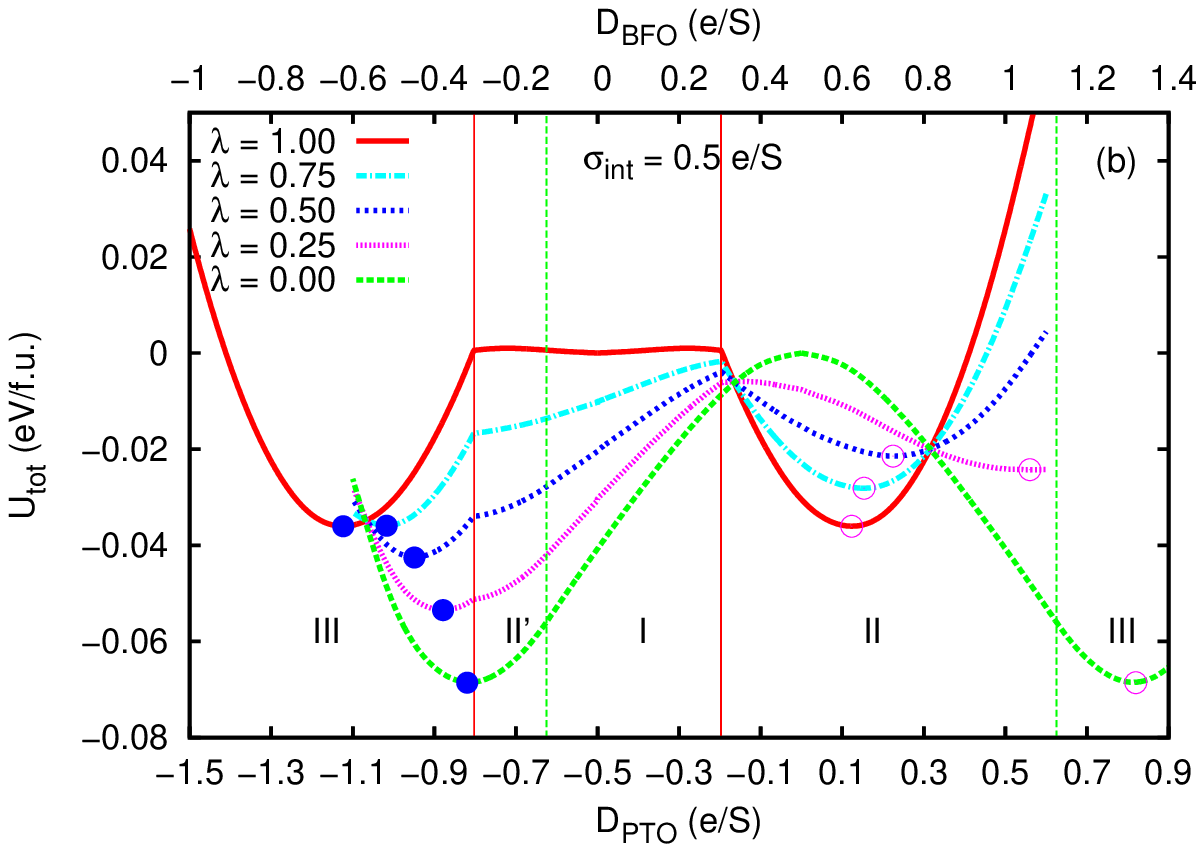}}%
\vspace{-0.30cm}
\caption{(Color online) Energy of PTO/BFO superlattices with $a = 3.81$~\AA~ expressed 
         as a function of $D$, for selected values of $\lambda$ and $\sigma_{\rm int}$.  
         Equilibrium and metastable superlattice states are represented with solid 
         and empty dots. Red (green) vertical lines indicate phase transitions occurring 
         in bulk BFO (PTO) under different $D$ conditions. (a) and (b) represent
         the cases of neutral and polar interfaces, respectively.}
\label{fig2}
\end{figure}

We start by expressing the total energy of a monodomain 
two-color superlattice (i.e., composed of species A and B) as, 
\begin{equation}  
U_{\rm tot} (D, \lambda, a) = \lambda \cdot U_{\rm A} (D, a) +
                     \left( 1 - \lambda \right) \cdot U_{\rm B} (D, a)~.  
\label{eq:totalenosig}
\end{equation}
Here $U_{\rm A}$ and $U_{\rm B}$ are the internal energies of the individual 
constituents, $D$ is the electric displacement along the out-of-plane 
stacking direction (i.e., $D \equiv {\cal E} + 4 \pi P$ where ${\cal E}$ is the 
electric field and $P$ is the \emph{effective} polarization, relative to the 
centrosymmetric reference configuration), $\lambda$ is the relative volume ratio of material A 
(i.e, $\lambda \equiv m / (n+m)$ where $m$ and $n$ are the thicknessess of layers
A and B, respectively), and $a$ is the in-plane lattice parameter (we assume 
heterostructures that are coherently strained to the substrate).
Note that short-range interface effects have been neglected. (While it is certainly
possible to incorporate the latter in the model, e.g. along the guidelines 
described in Ref.~\cite{wu08}, we believe these would have been
an unnecessary complication in the context of the present study.) 
By construction, Eq.~(\ref{eq:totalenosig}) implicitly enforces the continuity of $D$ 
along the out-of-plane stacking direction (which we label as $z$ henceforth), which is
appropriate for superlattices where the interfaces are nominally 
uncharged~[\onlinecite{ghosez08,junquera11}].

In presence of a polar mismatch, one has a net ``external'' 
interface charge, of compositional origin~\cite{murray09}, 
$\sigma_{\rm int}$ (see Fig.~\ref{fig1}a),
which is localized at the interlayer junctions.
In such a case, Eq.~(\ref{eq:totalenosig}) needs to be revised as follows,
\begin{equation}  
U_{\rm tot} (D, \sigma_{\rm int}, \lambda, a) = \lambda \cdot U_{\rm A} (D, a) +
            \left( 1 - \lambda \right) \cdot U_{\rm B} (D-\sigma_{\rm int}, a)~,  
\label{eq:totalen}
\end{equation}
i.e. the $U_{\rm B}$ curve is shifted in $D$-space to account for the jump in
$D$ produced by $\sigma_{\rm int}$. (Recall the macroscopic Maxwell equation,
$\nabla \cdot {\bf D} = \rho_{\rm ext}$, where $\rho_{\rm ext}$, the
``external'' charge, encompasses all contributions of neither dielectric nor 
ferroelectric origin.)
Once the functions $U_{\rm A}$ and $U_{\rm B}$ are computed and
stored (e.g. by using the methodology of Ref.~\cite{stengel09b}),
one can predict the ground-state of a hypothetical A/B superlattice  
by simply finding the global minimum of $U_{\rm tot}$ with respect 
to $D$ at fixed values of $\sigma_{\rm int}$, $\lambda$ and $a$.
The advantage of this procedure is that, for a given choice of A and B, the 
aforementioned four-dimensional parameter space can be explored 
very efficiently, as no further \emph{ab initio} calculations are
needed.

It is useful, before going any further, to specify the physical origin of 
$\sigma_{\rm int}$ in the context of this work. Consider, for example, 
a periodic BiFeO$_3$/PbTiO$_3$ superlattice, which we assume (i)~to be 
stoichiometric (and therefore charge-neutral) as a whole, (ii)~to have an ideal
AO-BO$_2$-AO-BO$_2$ stacking along the (001) direction, and (iii)~to form (say) 
AO-type interfaces (see Fig.~\ref{fig1}b). (The same arguments can be equally well 
applied to the case of BO$_2$-type interfaces.) Depending on the growth conditions, 
one can have a certain degree of intermixing in the boundary AO layers, which
will adopt an intermediate composition Bi$_x$ Pb$_{(1-x)}$O. As a pure
BiO layer is formally charged $+1$ and PbO is neutral, we can readily 
write $\sigma_{\rm int} = \pm \left(x-\frac{1}{2}\right)$ (expressed in 
units of $e/S$ with $S$ being the surface of the corresponding 5-atom perovskite cell), 
where the choice of plus or minus depends on the arbitrary assignment of 
BiFeO$_3$ and PbTiO$_3$ as the A or B component in Eq.~(\ref{eq:totalen}) 
[see Fig.~\ref{fig1}b]. 
In the following we shall illustrate the crucial role played by $\sigma_{\rm int}$
(and hence, by the interface stoichiometry) on the ferroelectric properties 
of a BFO/PTO superlattice, by combining Eq.~(\ref{eq:totalen}) with
the bulk $U_{\rm BFO} (D,a)$ and $U_{\rm PTO} (D,a)$ curves that we calculate 
from first principles.

Our calculations are performed with the ``in-house'' LAUTREC code 
within the local spin density approximation to density-functional theory.
(We additionally apply a Hubbard $U = 3.8$~eV to Fe ions~\cite{kornev07,yang12}.)
We use the $20$-atom simulation cell depicted in Fig.~\ref{fig1}c for both
BFO and PTO, which allows us to describe the 
ferroelectric and anti-ferrodistortive (AFD) modes of interest
(i.e. in-phase AFD$_{zi}$ and out-of-phase AFD$_{zo}$ and 
AFD$_{xy}$, see Ref.~[\onlinecite{bousquet08}]).
Atomic and cell relaxations are performed by constraining the out-of-plane
component of $D$~\cite{stengel09b} and the in-plane lattice constant $a$ to 
a given value. [Calculations are repeated many times in order to 
span the physically relevant two-dimensional $(D,a)$ parameter space.]

We start by illustrating the results obtained at fixed strain, $a = 3.81$~\AA~ 
(see Fig.~\ref{fig2}), by assuming $\sigma_{\rm int} = 0$, which corresponds to 
fully intermixed junctions ($x=0.5$), and we vary the BFO volume ratio, $\lambda$.
At the extreme values of $\lambda$, the results are consistent with the expectations: 
the equilibrium configuration of BFO (i.e., the minimum of $U_{tot}$ with $\lambda = 1$)
at this value of $a$ is the well-known R-type $Cc$-I phase~\cite{alison10}, 
derived from the bulk ground state via the application of epitaxial compression;
PTO ($\lambda = 0$), on the other hand, is in a tetragonal $P4mm$ phase with the 
polarization vector oriented out of plane.
Intermediate values of $\lambda$ yield a linear combination of the two 
single-component $U(D)$ curves, where the spontaneous $P_{z}$ at equilibrium 
gradually moves from the pure PTO to the pure BFO value.

Unfortunately, the possible equilibrium states that can be attained by 
solely varying $\lambda$ (at this value of $a$ and $\sigma_{\rm int}$)
lie far from any physically ``interesting'' region of the phase diagram.
For example, note the kink at $|D|\sim$0.3~C/m$^2$ in the pure BFO case, which 
corresponds to a first-order transition to an orthorhombic $Pna2_{1}$ phase
(a close relative of the higher-symmetry $Pnma$ phase, occurring at $D=0$). 
A huge piezoelectric and dielectric response
is expected in BFO in a vicinity of the transition~\cite{cazorla14}, raising
the question of whether one could approach this region by playing with $\sigma_{\rm int}$,
in addition to $\lambda$.

The answer is yes: when oxide superlattices with $\sigma_{\rm int} = 0.5$ are considered 
[corresponding to ``ideal'' (BiO)$^{+}$/TiO$_{2}$ and (FeO$_{2}$)$^{-}$/PbO interfaces],
the stable minimum of the system favors a smaller spontaneous polarization in the BFO
layers, approaching the aforementioned ($Cc{\rm -I} \to Pna2_{1}$) phase boundary in 
the limit of small $\lambda$.
Interestingly, the $U_{\rm tot}(D)$ curve becomes asymmetric (the interfacial charge breaks 
inversion symmetry), and a secondary, metastable minimum appears.
Overall, the resulting phase diagram turns out to be much richer, with new combinations 
of phases emerging (e.g. in region II', where 
BFO exists in the orthorhombic $Pna2_{1}$ phase and PTO in the 
tetragonal $P4mm$ phase), and highly non-trivial changes in the electrical 
properties occurring as a function of $\lambda$.

\begin{figure}
\centerline
        {\includegraphics[width=1.00\linewidth]{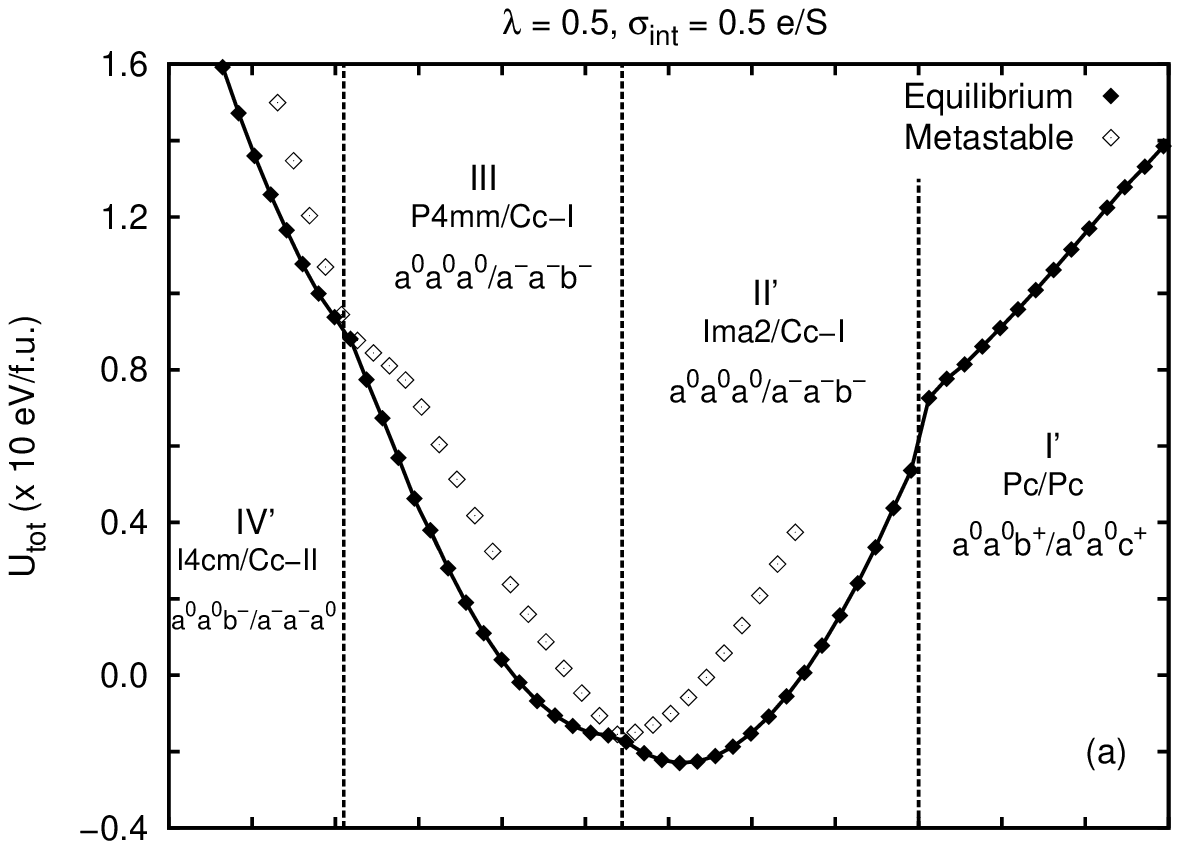}}%
        {\includegraphics[width=1.00\linewidth]{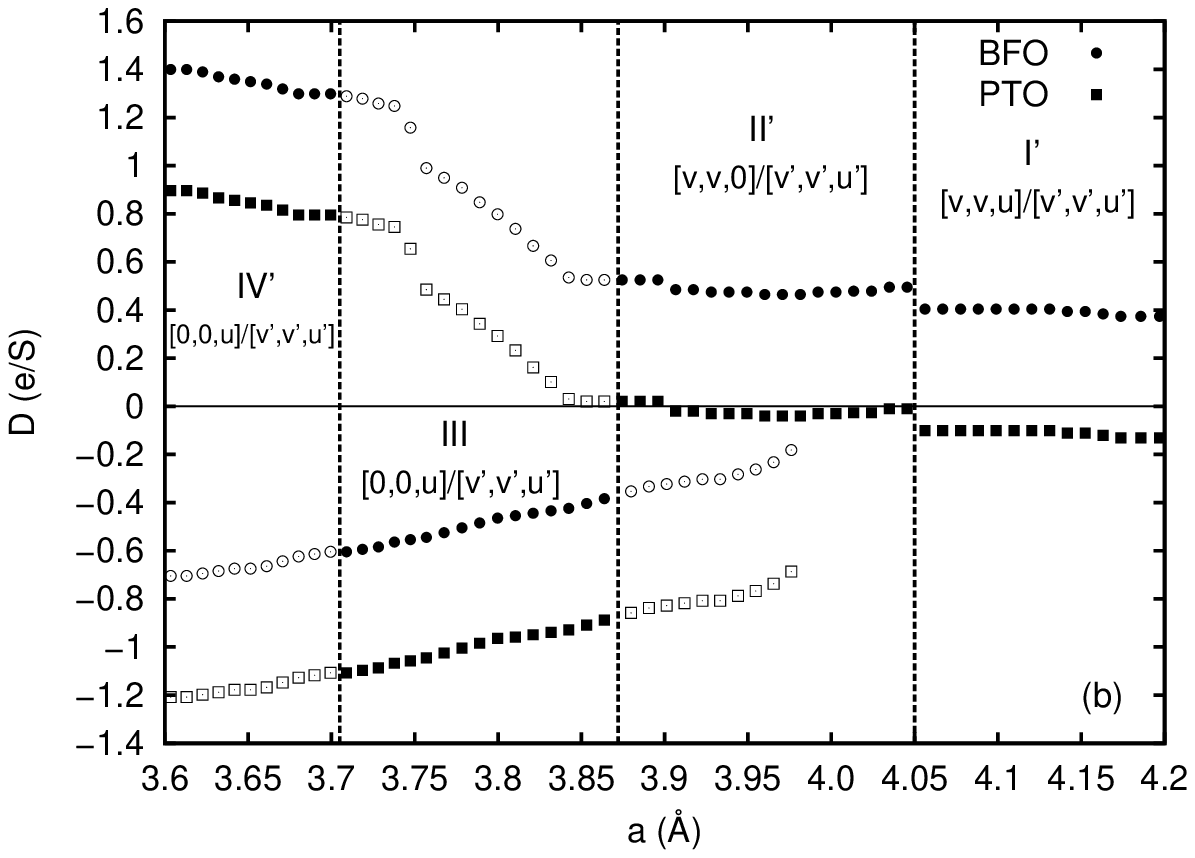}}%
\vspace{-0.30cm}
\caption{Total energy (a) and out-of-plane electric displacement $D$ (b) of
         the equilibrium (solid symbols) and metastable (empty symbols) states 
         of PTO/BFO superlattices with $\lambda = \frac{1}{2}$ and 
         $\sigma_{\rm int} = 0.5$, expressed as a function of the in-plane lattice 
         parameter. Regions in which PTO and BFO exist in different phases 
         are delimited with vertical dashed lines; the corresponding space groups 
         and AFD distortion patterns in Glazer's notation are shown in (a), 
         and the components of the ferroelectric polizarization in (b).}
\label{fig3}
\end{figure}

In order to further illustrate the power of our approach, we shall now 
fix the volume ratio to $\lambda=0.5$ (corresponding to alternating
BFO and PTO layers of equal thickness) and vary the in-plane lattice
parameter in the range $3.6 \le a \le 4.2$~\AA~.
We shall first consider the case of charged interfaces with $\sigma_{\rm int} = 0.5$,
as this choice allows for a direct comparison with the results of Yang \emph{et al.} 
(obtained via standard supercell simulations)~[\onlinecite{stengel12}].
In Fig.~\ref{fig3} we show the energy and spontaneous electric displacement of
the equilibrium and metastable states as a function of $a$. 
Four regions can be identified in the diagrams depending on the 
phases adopted by BFO and PTO at each value of the in-plane strain. 
(Their crystal space groups, AFD pattern and in-plane / out-of-plane ferroelectric polarization,
respectively $P_{xy}$ and $P_z$, are specified in compact form in the figure.)
In region I' both PTO and BFO adopt a monoclinic $Pc$ phase characterized 
by large in-phase AFD$_z$ distortions and non-zero $P_{xy}$ and $P_z$. 
Such a monoclinic $Pc$ phase is closely related to the orthorhombic $Pmc2_{1}$ 
structure which has been recently predicted in PTO and BFO at large tensile 
strains~[\onlinecite{yang12}]. 
In region II' PTO adopts an orthorhombic $Ima2$ phase, characterized 
by vanishing AFD distortions and a large in-plane ${\bf P}$ (we neglect
the small out-of-plane $P_z$), while BFO is in its well-known $Cc$-I state. 
In region III, BFO remains $Cc$-I, while PTO adopts a $P4mm$ phase, both 
with \emph{opposite} out-of-plane polarization with respect to region II'.
These structures switch back to a positively oriented $P_z$ in region IV',
respectively transforming into a monoclinic $Cc$-II and a tetragonal $I4cm$ phase. 
The $I4cm$ phase is characterized by anti-phase AFD$_z$ distorsions and an out-of-plane ${\bf P}$, 
while the $Cc$-II corresponds to the ``supertetragonal'' T-type phase of BFO~\cite{zeches09}.
Note that, as observed already while discussing Fig.~\ref{fig2}, the net interface
charge leads to an asymmetric double-well potential, and consequently to an energy
difference (typically of $\sim 20$~meV/f.u. or less, see Fig.~\ref{fig3}a) between 
the two oppositely polarized states. (Only one minimum survives at large tensile
strains, where the superlattice is no longer ferroelectric.)
At the phase boundaries such energy difference vanishes;
the obvious kinks in the $U_{\rm tot}$ curve shown in Fig.~\ref{fig3}(a) 
indicate that the transitions (at $a = 3.71$, $3.87$, and $4.05$~\AA) are all 
of first-order type. 

The above results are in remarkable agreement with those of  Yang \emph{et al.}~\cite{stengel12}.
The only apparent discrepancy concerns the ordering of the stable/metastable states in
region III, which anyway involves a very subtle energy difference (and is therefore sensitive
to short-range interface effects, not considered here). 
Obtaining such an accurate description of superlattices where the individual layers are
as thin as three perovskite units~\cite{stengel12} provides a stringent benchmark for
our method, and validates it as a reliable modeling tool.
From the physical point of view this comparison suggest that, even in the ultrathin limit, 
PTO/BFO superlattices can be well understood in terms of macroscopic bulk effects, i.e.,
short-range interface-specific phenomena appear to play a relatively minor role.

\begin{figure}
\centerline
        {\includegraphics[width=1.00\linewidth]{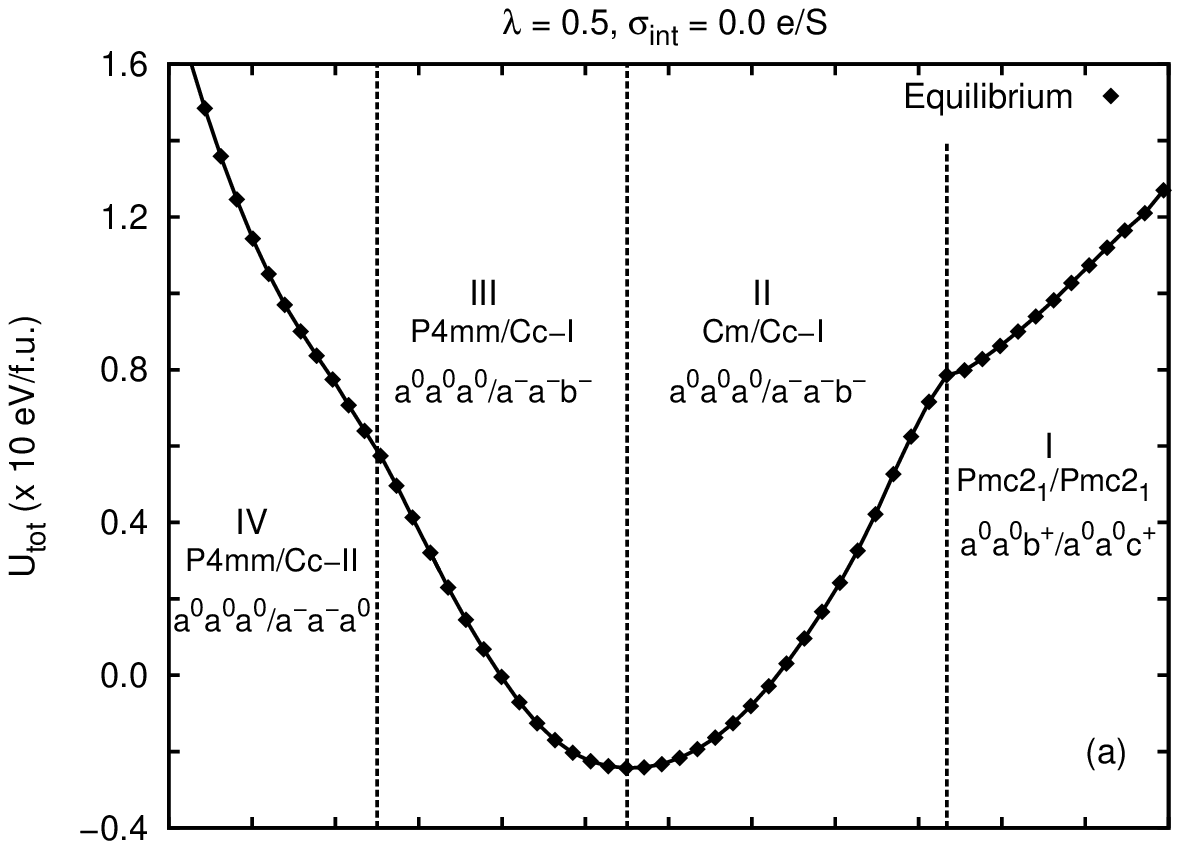}}%
        {\includegraphics[width=1.00\linewidth]{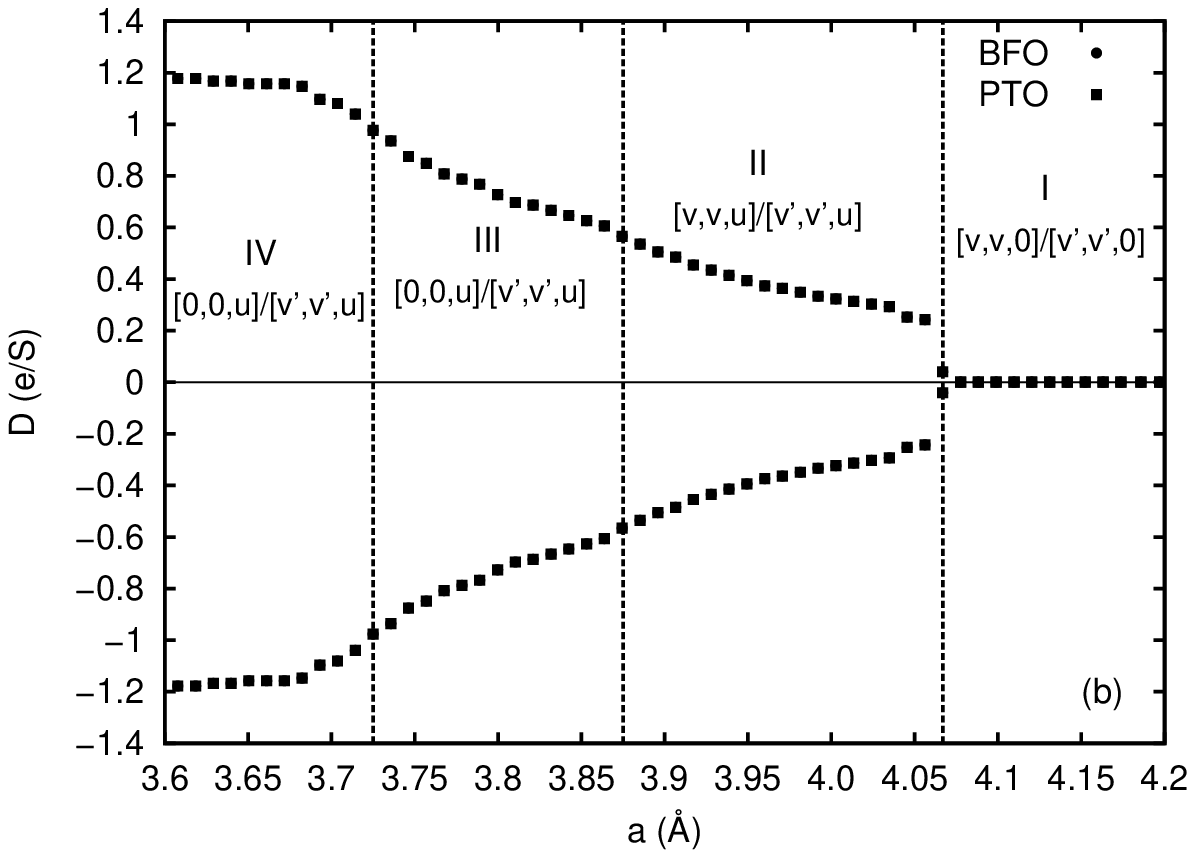}}%
\vspace{-0.30cm}
\caption{Same as in Fig.~\ref{fig3}, but considering neutral
         interfaces. The out-of-plane polarization is the 
         same in PTO and BFO layers.}
\label{fig4}
\end{figure}

Having gained confidence in our method, we can use it to predict the behavior of
a hypothetical superlattice with $\sigma_{\rm int} = 0$, corresponding to a 
centrosymmetric reference structure with fully intermixed Pb$_{0.5}$Bi$_{0.5}$O 
interface layers (see Fig.~\ref{fig4}).
Note the symmetry of the two opposite polarization states, and the common value of
the spontaneous electric displacement adopted by BFO and PTO.
The resulting phase diagram consists, again, in four regions, with 
a first-order and two second-order phase transitions occurring at $a = 4.07$, 
$3.88$ and $3.73$~\AA~, respectively (see Fig.~\ref{fig4}a). 
In three of these regions, the individual layers display structures which are 
different from those obtained in the $\sigma_{\rm int}= 0.5$ case: in region I 
both PTO and BFO stabilize in an orthorhombic $Pmc2_{1}$ phase~[\onlinecite{yang12}],
characterized by a vanishing $P_z$;
in region II PTO adopts a monoclinic $Cm$ phase with
the polarization roughly oriented along (111) ($P_z \neq P_{xy} \neq 0$) and no AFD, 
while BFO stabilizes in the already discussed $Cc$-I phase;
finally, in region IV, PTO is tetragonal $P4mm$ and BFO is monoclinic $Cc$-II.
These findings quantitatively demonstrates that the interface 
charge mismatch can have a tremendous impact on the physical properties of 
oxide superlattices. Our simple and general method allows one to model and quantify 
accurately these effects, and most importantly to rationalize them in 
terms of intuitive physical concepts.

In summary, we have discussed a general theoretical framework to predict
the behavior of charge-mismatched superlattices. We have showed that the
effect of the interface stoichiometry, which we describe via the 
``compositional'' interface charge $\sigma_{\rm int}$, is quite dramatic,
and needs to be properly accounted for in the models.
More generally, we argue that $\sigma_{\rm int}$ can be regarded, in addition 
to $\lambda$ and $a$, as a further degree of freedom in designing
oxide heterostructures with tailored functionalities, opening
exciting new avenues for future research.

\acknowledgments
This work was supported by MICINN-Spain [Grants No. MAT2010-18113 and
No. CSD2007-00041], and the CSIC JAE-DOC program (C.C.). We 
thankfully acknowledge the computer resources, technical expertise and
assistance provided by RES and CESGA.

\end{document}